\documentclass[10pt,twocolumn]{article}
\usepackage{cite}

\newcommand{\tsc}[1]{\textsc{#1}}
\newcommand{\mrm}[1]{\mathrm{#1}}

\newcommand{\B}{\ensuremath{\mrm{B}}}
\newcommand{\Z}{\ensuremath{\mrm{Z}}}
\newcommand{\brm}{\ensuremath{\mrm{b}}}

\newcommand{\s}{\ensuremath{\mrm{s}}}

\def\lsim{\mathrel{\rlap{\lower2pt\hbox{\hskip1pt$\sim$}}
\raise2pt\hbox{$<$}}}
\def\gsim{\mathrel{\rlap{\lower2pt\hbox{\hskip1pt$\sim$}}
\raise2pt\hbox{$>$}}}

\begin{document}
\title{A Quick Guide to SUSY
  Tools\footnote{FERMILAB-CONF-06-004-T. Prepared for 
  the TeV4LHC Physics Landscapes Summary Report.}}
\author{Peter Z.\ Skands\footnote{skands@fnal.gov}\\\em\small Theoretical Physics, Fermi National Accelerator Laboratory, Batavia, IL, USA}
\date{}
\maketitle
\begin{abstract}
The last decade has seen the emergence of 
a wide range of automated calculations for supersymmetric extensions
of the Standard Model. This guide contains a brief summary of these,
with the main focus on hadron collider phenomenology, 
as well as a brief introduction to the so-called SUSY Les Houches
Accord. See also the Les Houches Web Repository for BSM Tools:\\
\texttt{http://www.ippp.dur.ac.uk/montecarlo/BSM/}  
\end{abstract}

\section{Introduction}
Among the most enticing possibilities for observable New Physics both at
the Tevatron and at the Large Hadron Collider is supersymmetry
(SUSY); for reviews, see e.g.\ \cite{Martin:1997ns,Tata:1997uf,Chung:2003fi}. 
At the most fundamental level, imposing supersymmetry on a quantum
field theory represents the most general (and only) 
possible way of extending the Poincar\'e group of space--time
symmetries \cite{coleman67,haag75}, 
at the same time as it furnishes a desirable relation between the
bosonic and fermionic degrees of freedom. Empirically, however, 
supersymmetry can at most be a broken symmetry if it exists in
Nature, due to the non-observation of a mass-degenerate (or lighter) 
spin-partner for each of the Standard Model (SM) particles. 

However, even a softly broken supersymmetry can have 
quite amazing properties, as long as the mass
splittings introduced by the breaking are smaller than a
TeV or so. 
Among the most well-known consequences of such a type of
supersymmetry are radiative breaking of electroweak symmetry, 
an elegant solution to the so-called 
hierarchy problem, a natural weakly interacting dark matter candidate
(in theories with conserved $R$-parity), and unification of the
strong, weak, and electromagnetic gauge couplings at a (very) 
high energy scale. 

For collider phenomenology, the most immediately relevant
consequences are 1) an extension of the
Standard Model Higgs sector to (at least) 2 doublets, 2)
promotion of each of the Standard Model fields (plus the extra Higgs
content) to superfields, resulting in a spin-partner for each SM particle, with
mass splittings inside each boson-fermion doublet $\lsim$ 1 TeV, and
3) the special properties which accompany a conserved $R$-parity,
namely production of the new states only by the pair, followed by
individual cascade decays down to the Lightest Supersymmetric
Particle (LSP) which is stable and (usually) escapes detection.

The large interest in ($N=1$) supersymmetric extensions of the SM and 
their phenomenological consequences has carried
with it the need for automated tools to calculate supersymmetric mass
and coupling spectra, cross sections, decay rates, dark matter relic
densities, event rates, precision observables, etc. 
To handle the cross-communication between
the many tools,
the so-called SUSY Les Houches Accord
\cite{Skands:2003cj,Allanach:2004ub,slha2} (SLHA)  
is now in widespread use. Section \ref{sec:slha} contains a brief
introduction to this Accord. Next, in Section \ref{sec:tools}, 
an overview of the presently available state-of-the-art tools is
given, divided into four main categories. A more extensive collection
of tools for BSM physics as well as an online repository can be found
in \cite{bsmrepository}. Another recent and comprehensive 
tools review is the Les Houches Guidebook to MC Generators
\cite{Dobbs:2004qw}. 

\section{The SUSY Les Houches Accord \label{sec:slha}}
Given the long history of the subject, it is not surprising that
several different conventions for defining supersymmetric theories
have been proposed over the years, many of which are in active use 
by different groups. While this is not a problem per se (unique
translations can always be constructed), it does entail a few
practical problems, in particular when the results of one group
are compared to (or used for) the calculations of a different group. 

In addition, even when the theoretical conventions are identical, there
remains the purely technical issue that each program has its own native way
of inputting and outputting parameters, each of which is
unintelligible to most other programs.  

The SLHA was proposed to solve both these problems. Due to the large
parameter space of unconstrained supersymmetric models, the SLHA in
its present form \cite{Skands:2003cj} (SLHA1)
is limited to the MSSM, with conservation of $R$-parity, CP,
and flavour. Extensions to more general models are underway
\cite{slha2} (SLHA2). 

Technically, the Accord is structured into 3 ASCII files 
(or strings): 1) model
definition and measured SM parameters, 2) SUSY mass and coupling
spectrum, and 3) decay tables. Though admittedly not elegant, the
ASCII format was chosen for its robustness across platforms and
compilers. In general, all input parameters used for a calculation 
are copied to the output, so that any subsequent calculation also has access
to the exact input parameters used for the previous one. 

\subsection{The SLHA Conventions}
The backbone of the Accord is a unique set of conventions for defining
the supersymmetric parameters, fields, and couplings. These
conventions, which have also been adapted for the so-called SPA
project \cite{Aguilar-Saavedra:2005pw}, 
largely resemble the widely used Gunion-Haber conventions
\cite{Gunion:1986yn}, with a few differences as noted explicitly in
\cite{Skands:2003cj}. Simply stated, 
to define a SUSY model, one needs the field content, the 
Superpotential, the SUSY breaking terms, and the gauge couplings. For
the field content, the SLHA assumes that of the MSSM, while SLHA2 will
include extensions for the NMSSM. 

The MSSM Superpotential is 
specified by the measured SM particle masses (giving the Yukawa
couplings) and by the $\mu$ term. At present, only the third
generation Yukawas are included. The gauge couplings are specified in
terms of $M_\Z$, $G_F$, $\alpha_s(M_\Z)^{\overline{\mrm{MS}}}$, 
and the fine structure constant at zero momentum transfer. All of
these are the standard SM ones that one can get
from a review text, i.e.\ no SUSY corrections should be
included here. SLHA2 will include masses for all 3 generations, as
well as the CKM matrix.

The SUSY breaking terms can either be specified by giving the
parameters for a minimal version of a particular SUSY breaking model
(SUGRA, GMSB, or AMSB), or individually, either by
starting from a minimal model and successively adding non-universal
terms or by simply giving all terms explicitly. 
For higher-order calculations, 
these parameters are interpreted as given in the modified dimensional
reduction ($\overline{\mrm{DR}}$) scheme
\cite{Siegel:1979wq,Capper:1980ns,Jack:1994rk}, either at the 
(derived) unification scale or at a 
user-specifiable scale. As mentioned, CP, $R$-parity, and flavour are
assumed conserved in SLHA1.

In the spectrum output, three kinds of parameters are given: 1) pole
masses of all (s)particles, 2) mixing matrices, and 3) Lagrangian
parameters. While the precise definition of the mixing matrix elements are left
up to each spectrum calculator, the Lagrangian parameters are
defined as $\overline{\mrm{DR}}$ ones at one or several
user-specifiable scales $Q$.

\subsection{The SLHA Decay Tables}
A somewhat separate and self-contained aspect of the SLHA is the
possibility to pass total widths and partial branching ratios via a
file structure similar to that of the rest of the Accord. A common use
for this is to improve or extend the width calculations of an
event generator by the numbers calculated with a specialised
package. 

{\bf Note!} An important potential pitfall when using these files is 
on-shell intermediate resonances in final states
with more than 2 particles. If not treated properly, 
large problems both with double-counting and with incorrect population of
phase space can occur. Please see \cite{Skands:2003cj} for an
explicit description of the correct procedure to adopt in these cases.

\section{Computing SUSY \label{sec:tools}}
This Section contains an overview of SUSY calculational tools, divided
into 1) spectrum calculators, 2) observables calculators, 3) matrix
element and event generators, and 4) fitting programs. For links and
references, the reader should consult the recently constructed online
repository for BSM tools \cite{bsmrepository}. 

\subsection{Spectra}
Given assumptions about the
underlying supersymmetric theory (field content, superpotential, supersymmetry
breaking terms) and a set of measured parameters (SM particle masses,
gauge couplings, electroweak symmetry breaking), the masses and
couplings of all particles in the spectrum can be computed. 
This is the task of spectrum calculators, also called RGE packages. 

The most commonly used all-purpose spectrum calculators are
\tsc{Isajet} \cite{Baer:1993ae},
\tsc{SoftSusy} \cite{Allanach:2001kg}, \tsc{SPheno} \cite{Porod:2003um}, and
\tsc{SuSpect} \cite{Djouadi:2002ze}, all compatible with
SLHA. In general, the codes agree with each other to within a percent
or so, though larger discrepancies can occur in particular at large
$\tan\beta$. For mSUGRA, a useful online tool for comparison between
them (and different versions of them)
exists \cite{kraml}. Recent detailed comparison studies can be found in
\cite{Allanach:2003jw,Allanach:2004rh,Belanger:2005jk}. 
Though \tsc{Pythia} also contains an internal spectrum
calculator \cite{Mrenna:1996hu}, 
the resulting spectrum is very approximate 
and should not be used for serious studies. 

There are also a few spectrum calculators with more specialised areas
of application, here \tsc{CPSuperH} \cite{Lee:2003nt}, \tsc{FeynHiggs}
\cite{Heinemeyer:1998yj}, and 
\tsc{NMHDecay} \cite{Ellwanger:2004xm}. 
\tsc{NMHDecay} computes the entire mass spectrum in
the NMSSM (and has a limit which is equivalent to the MSSM), 
but couplings and decay widths are so far only calculated for the Higgs
sector, though improvements are underway. \tsc{NMHDecay} is compatible
with an extension of the SLHA \cite{slha2}. 
The program \tsc{FeynHiggs} deals with
the Higgs sector of the MSSM, for which it contains higher
precision calculations than the general-purpose programs mentioned
above. It is also able to handle both minimal flavour violation (MFV) and CP
violation, and is compatible with the SLHA, hence can e.g.\ be used to
provide a final adjustment to the Higgs sector of a general spectrum
calculated by one of the other codes. Finally, \tsc{CPSuperH} deals
with the Higgs sector in the MSSM with explicit CP violation and
contains a number of refinements which makes it interesting also in the
CP conserving case. 

\subsection{Observables}
Programs 
that calculate one or more of the following: cross sections,
decay partial widths, dark matter relic density, and
indirect/precision observables. Note that we here focus on 
calculations relevant for hadron colliders and that 
matrix element and event generators, which
also calculate many of these things, are treated separately below.

For hadron collider cross sections, \tsc{Prospino} \cite{Beenakker:1996ed}
can be used to calculate inclusive SUSY-NLO cross sections, both total
and differential. It also calculates the LO cross section and gives
the corresponding K-factor. 

For decay partial widths, several specialised packages exist.
For the MSSM, \tsc{SPheno} calculates tree-level
decays of all (s)particles (soon to include RPV\footnote{RPV in SPheno
  is not yet public, but a private version 
  is available from the author}), 
\tsc{SDecay} \cite{Muhlleitner:2003vg} 
computes sparticle decay widths including NLO SUSY-QCD
effects, and both 
\tsc{FeynHiggs} \cite{Heinemeyer:1998yj} 
and \tsc{HDecay} \cite{Djouadi:1998yw} 
compute Higgs partial widths with higher-order
corrections. \tsc{NMHDecay} \cite{Ellwanger:2004xm} computes
partial widths for all Higgs bosons in the NMSSM.

For the density of dark matter, \tsc{DarkSUSY} \cite{Gondolo:2004sc},
\tsc{IsaTools} \cite{Baer:2003jb},  
and \tsc{MicrOMEGAs} \cite{Belanger:2001fz}
represent the publically available state-of-the-art tools. All of
these work for the MSSM, though a special effort has been put into 
\tsc{MicrOMEGAs} to make it easily extendable \cite{micromegas2},
recently resulting in 
an implementation of the NMSSM \cite{Belanger:2005kh}, and work
on CP violation is in progress.

For precision observables, \tsc{micrOMEGAs} includes calculations of
$(g-2)_\mu$, $\brm\to\s\gamma$, $\B_\s\to\mu^+\mu^-$, and cross-sections
for neutralino annihilation at v$\sim$0, relevant for indirect detection of
neutralinos. 
\tsc{NMHDecay} includes a check against LEP
Higgs searches, $\brm\to\s\gamma$, and can be interfaced to
\tsc{MicrOMEGAs} for the relic density. \tsc{Isajet}/\tsc{IsaTools} 
include calculations of $\brm\to\s\gamma$, $(g-2)_\mu$,
$\B_\s\to\mu^+\mu^-$, $\B_d\to\tau^+\tau^-$, 
and neutralino-nucleon scattering cross sections. 
\tsc{SPheno} includes $\brm\to\s\gamma$,  $(g-2)_\mu$, as well as the
SUSY contributions to the $\rho$ parameter due to sfermions. 
\tsc{FeynHiggs} also evaluates the contribution to
electroweak precision observables via $\Delta\rho$ as well as $(g-2)_\mu$
(with two-loop corrections). 
Finally, \tsc{SuSpect} also includes a calculation of $\brm\to\s\gamma$.

\subsection{Matrix Element and Event Generators}
By a matrix element generator, we here understand a program that,
given a set of fields and a Lagrangian, is able to generate Feynman
diagrams for any process and square them. Note, however, that many of
the codes are able to do quite a bit more than that. An event
generator is a program that, given a matrix element, is able to
generate a series of random exclusive events in phase space, often
including resonance decays, parton showers, underlying
event, hadronisation, and hadron decays.

The automated tools for generating matrix
elements for SUSY are \tsc{Amegic++} \cite{Krauss:2001iv}, \tsc{CalcHEP}
\cite{Pukhov:2004ca}, \tsc{CompHEP} \cite{Pukhov:1999gg},
\tsc{Grace-SUSY} \cite{Tanaka:1997qn}, 
\tsc{SUSY-MadGraph} \cite{Cho:2006sx}, and \tsc{O'Mega}
\cite{Moretti:2001zz}.  All of these work
at Leading Order, except \tsc{Grace}, and all currently 
only deal with the MSSM, except
\tsc{CalcHEP} which contains an NMSSM implementation. 

\tsc{CalcHEP} and \tsc{CompHEP} provide internal event generators, 
while the event generator \tsc{Sherpa} \cite{Gleisberg:2003xi} is built on
\tsc{Amegic++},  
\tsc{Gr@ppa} \cite{Tsuno:2005ih} builds on
\tsc{Grace}, \tsc{MadEvent} \cite{Maltoni:2002qb} builds on
\tsc{MadGraph} (work is in progress to extend this to 
\tsc{SUSY-MadGraph}), and
\tsc{Whizard} \cite{Kilian:2001qz} builds on \tsc{O'Mega}. Of these, most are 
matrix-element-level event generators, that is they
provide events consisting of just a few partons and their four-momenta, 
corresponding to the given matrix element convoluted with phase
space. These events must then be externally interfaced \cite{Boos:2001cv} 
e.g.\ to \tsc{Pythia} or
\tsc{Herwig} for resonance decays, parton showering, underlying event, and
hadronisation. The exception is \tsc{Sherpa}, which contains its own
parton showers and underlying-event models 
(similar to the \tsc{Pythia} ones), and for which a
cluster-based hadronisation model is being developed. 

In addition, both \tsc{Herwig} \cite{Corcella:2000bw} and \tsc{Pythia} 
\cite{Sjostrand:2000wi} contain a large number of
internal hardcoded leading order matrix elements, 
including $R$-parity violating (RPV) decays in
both cases
\cite{Dreiner:1999qz,Richardson:2000nt,Skands:2001it,Skands:2002qz,Sjostrand:2002ip}, 
and RPV single sparticle production in
\tsc{Herwig} \cite{Dreiner:1999qz}. In \tsc{Pythia}, the parton shower
off SUSY resonance decays is merged to the real NLO jet emission matrix
elements \cite{Norrbin:2000uu}, 
an interface to \tsc{CalcHEP} and \tsc{NMHDecay} exists
for the NMSSM \cite{Pukhov:2005je}, and an implementation of 
the hadronisation of $R$-hadrons is available
\cite{Kraan:2004tz,Kraan:2005ji}.  

Two other event generators should be mentioned. \tsc{Isajet}
\cite{Baer:1993ae} also
contains a large amount of SUSY phenomenology, but 
its parton shower and hadronisation machinery 
are much less sophisticated than those of \tsc{Herwig}, 
\tsc{Pythia}, and \tsc{Sherpa}. The active development of
\tsc{Susygen} \cite{Ghodbane:1999va} 
(which among other things 
includes RPV single sparticle production) 
is currently at a standstill, though basic maintenance is
still being carried out. 

\subsection{Fitters}
Roughly speaking, the tools described above all have one thing in
common: given a set of fundamental parameters
(themselves not directly observable) they calculate the
(observable) phenomenological and experimental consequences. However, 
if SUSY is at some point discovered, a somewhat complementary game will
ensue: given a set of observed masses, cross sections, and branching ratios, 
how much can we say about the fundamental parameters? 

The fitting programs \tsc{Fittino} \cite{Bechtle:2004pc} and \tsc{Sfitter} 
\cite{Lafaye:2004cn} attempt to address this question. 
In a spirit similar to codes like \tsc{Zfitter} \cite{Bardin:1999yd}, 
they combine the above tools in an automated statistical analysis, 
taking as input a set of measured observables and yielding as output a
set of fundamental parameters.

Obviously, the main difficulty does not lie in determining the actual central
values of the parameters, although this can require
significant computing resources in itself; by far the most
important aspect of these tools is the error analysis. Statistical
uncertainties can be treated rigorously, and are included in both
programs. Theoretical and systematic uncertainties
are more tricky. In a conventional analysis, these uncertainties are
evaluated by careful consideration of both
the experimental setup, and of the particular theoretical calculations
involved. In an automated analysis, which has to deal simultaneously 
with the entire parameter space of supersymmetry, a `correct' 
evaluation of these errors poses a truly formidable challenge, one
that cannot be considered fully dealt with yet. 

\section*{Acknowledgments}
The TeV4LHC series of workshops have been highly enjoyable, and I am
thankful for all the hard work put in by the organisers. Thanks also
to S.~Mrenna, W.~Porod, S.~Kraml, and S.~Heinemeyer 
for useful comments on the draft. This work was
supported by Universities Research Association 
Inc. under Contract No. DE-AC02-76CH03000 with the United States Department of
Energy.

\bibliography{skands_susytools_rev}

\begin{thebibliography}{10}

\bibitem{Martin:1997ns}
S.~P. Martin,
\newblock (1997),
\newblock hep-ph/9709356.
%%CITATION = HEP-PH 9709356;%%

\bibitem{Tata:1997uf}
X.~Tata,
\newblock (1997),
\newblock hep-ph/9706307.
%%CITATION = HEP-PH 9706307;%%

\bibitem{Chung:2003fi}
D.~J.~H. Chung{ et~al.},
\newblock Phys. Rept. {\bf 407}, 1 (2005).
%%CITATION = HEP-PH 0312378;%%

\bibitem{coleman67}
S.~Coleman, J.~Mandula,
\newblock Phys. Rev. {\bf 159}, 1251 (1967).
%%CITATION = PHRVA,159,1251;%%

\bibitem{haag75}
R.~Haag, J.~T. Lopuszanski, M.~Sohnius,
\newblock Nucl. Phys. {\bf B}, 257 (1975).
%%CITATION = NUPHA,B88,257;%%

\bibitem{Skands:2003cj}
P.~Skands{ et~al.},
\newblock JHEP {\bf 07}, 036 (2004),
\newblock (see {\texttt{http://home.fnal.gov/$\sim$skands/slha/}}).
%%CITATION = HEP-PH 0311123;%%

\bibitem{Allanach:2004ub}
Beyond the Standard Model Working Group, B.~C. Allanach{ et~al.},
\newblock (2004),
\newblock hep-ph/0402295.
%%CITATION = HEP-PH 0402295;%%

\bibitem{slha2}
B.~C. Allanach{ et~al.},
\newblock {Status of the SUSY Les Houches Accord II Project},
\newblock {FERMILAB-CONF-05-517-T}. In {Les Houches} {'Physics at TeV Colliders
  2005'} {BSM Working Group}: {Summary report}, hep-ph/0602198.

\bibitem{bsmrepository}
P.~Skands{ et~al.},
\newblock A repository for beyond-the-standard-model tools,
\newblock FERMILAB-CONF-05-521-T. In {Les Houches} {'Physics at TeV Colliders
  2005'} {BSM Working Group}: {Summary report}, hep-ph/0602198. See also:
  {\\\texttt{http://www.ippp.dur.ac.uk/montecarlo/BSM/}}.

\bibitem{Dobbs:2004qw}
M.~A. Dobbs{ et~al.},
\newblock (2004),
\newblock in Les Houches 'Physics at Tev Colliders 2003' QCD/SM Working Group:
  Summary Report (hep-ph/0403100). hep-ph/0403045.
%%CITATION = HEP-PH 0403045;%%

\bibitem{Aguilar-Saavedra:2005pw}
J.~A. Aguilar-Saavedra{ et~al.},
\newblock (2005),
\newblock hep-ph/0511344.
%%CITATION = HEP-PH 0511344;%%

\bibitem{Gunion:1986yn}
J.~F. Gunion, H.~E. Haber,
\newblock Nucl. Phys. {\bf B272}, 1 (1986),
\newblock [Erratum-ibid.\ {\bf B402}, 567 (1993)].
%%CITATION = NUPHA,B272,1;%%

\bibitem{Siegel:1979wq}
W.~Siegel,
\newblock Phys. Lett. {\bf B84}, 193 (1979).
%%CITATION = PHLTA,B84,193;%%

\bibitem{Capper:1980ns}
D.~M. Capper, D.~R.~T. Jones, P.~van Nieuwenhuizen,
\newblock Nucl. Phys. {\bf B167}, 479 (1980).
%%CITATION = NUPHA,B167,479;%%

\bibitem{Jack:1994rk}
I.~Jack, D.~R.~T. Jones, S.~P. Martin, M.~T. Vaughn, Y.~Yamada,
\newblock Phys. Rev. {\bf D50}, 5481 (1994).
%%CITATION = HEP-PH 9407291;%%

\bibitem{Baer:1993ae}
H.~Baer{ et~al.},
\newblock (1993),
\newblock hep-ph/9305342.
%%CITATION = HEP-PH 9305342;%%

\bibitem{Allanach:2001kg}
B.~C. Allanach,
\newblock Comput. Phys. Commun. {\bf 143}, 305 (2002).
%%CITATION = HEP-PH 0104145;%%

\bibitem{Porod:2003um}
W.~Porod,
\newblock Comput. Phys. Commun. {\bf 153}, 275 (2003).
%%CITATION = HEP-PH 0301101;%%

\bibitem{Djouadi:2002ze}
A.~Djouadi, J.-L. Kneur, G.~Moultaka,
\newblock (2002),
\newblock hep-ph/0211331.
%%CITATION = HEP-PH 0211331;%%

\bibitem{kraml}
S.~Kraml,
\newblock {\texttt{http://cern.ch/kraml/comparison/}}.

\bibitem{Allanach:2003jw}
B.~C. Allanach, S.~Kraml, W.~Porod,
\newblock JHEP {\bf 03}, 016 (2003).
%%CITATION = HEP-PH 0302102;%%

\bibitem{Allanach:2004rh}
B.~C. Allanach{ et~al.},
\newblock JHEP {\bf 09}, 044 (2004).
%%CITATION = HEP-PH 0406166;%%

\bibitem{Belanger:2005jk}
G.~B{\'e}langer, S.~Kraml, A.~Pukhov,
\newblock Phys. Rev. {\bf D72}, 015003 (2005).
%%CITATION = HEP-PH 0502079;%%

\bibitem{Mrenna:1996hu}
S.~Mrenna,
\newblock Comput. Phys. Commun. {\bf 101}, 232 (1997).
%%CITATION = HEP-PH 9609360;%%

\bibitem{Lee:2003nt}
J.~S. Lee{ et~al.},
\newblock Comput. Phys. Commun. {\bf 156}, 283 (2004).
%%CITATION = HEP-PH 0307377;%%

\bibitem{Heinemeyer:1998yj}
S.~Heinemeyer, W.~Hollik, G.~Weiglein,
\newblock Comput. Phys. Commun. {\bf 124}, 76 (2000),
\newblock see {\\\texttt{http://www.feynhiggs.de}}.
%%CITATION = HEP-PH 9812320;%%

\bibitem{Ellwanger:2004xm}
U.~Ellwanger, J.~F. Gunion, C.~Hugonie,
\newblock JHEP {\bf 02}, 066 (2005).
%%CITATION = HEP-PH 0406215;%%

\bibitem{Beenakker:1996ed}
W.~Beenakker, R.~Hopker, M.~Spira,
\newblock (1996),
\newblock hep-ph/9611232. See{\\ \tt
  http://pheno.physics.wisc.edu/$\sim$plehn/}.
%%CITATION = HEP-PH 9611232;%%

\bibitem{Muhlleitner:2003vg}
M.~M{\"u}hlleitner, A.~Djouadi, Y.~Mambrini,
\newblock (2003),
\newblock hep-ph/0311167.
%%CITATION = HEP-PH 0311167;%%

\bibitem{Djouadi:1998yw}
A.~Djouadi, J.~Kalinowski, M.~Spira,
\newblock Comput. Phys. Commun. {\bf 108}, 56 (1998).
%%CITATION = HEP-PH 9704448;%%

\bibitem{Gondolo:2004sc}
P.~Gondolo{ et~al.},
\newblock JCAP {\bf 0407}, 008 (2004).
%%CITATION = ASTRO-PH 0406204;%%

\bibitem{Baer:2003jb}
H.~Baer, C.~Balazs, A.~Belyaev, J.~O'Farrill,
\newblock JCAP {\bf 0309}, 007 (2003),
\newblock see {\\\tt http://www.phy.bnl.gov/$\sim$isajet/}.
%%CITATION = HEP-PH 0305191;%%

\bibitem{Belanger:2001fz}
G.~B{\'e}langer, F.~Boudjema, A.~Pukhov, A.~Semenov,
\newblock Comput. Phys. Commun. {\bf 149}, 103 (2002).
%%CITATION = HEP-PH 0112278;%%

\bibitem{micromegas2}
G.~B{\'e}langer, F.~Boudjema, A.~Pukhov, A.~Semenov,
\newblock {micrOMEGAS2.0} and the relic density of dark matter in a generic
  model,
\newblock In {Les Houches} {'Physics at TeV Colliders 2005'} {BSM Working
  Group}: {Summary report}, hep-ph/0602198.

\bibitem{Belanger:2005kh}
G.~B{\'e}langer{ et~al.},
\newblock JCAP {\bf 0509}, 001 (2005), [hep-ph/0505142].
%%CITATION = HEP-PH 0505142;%%

\bibitem{Krauss:2001iv}
F.~Krauss, R.~Kuhn, G.~Soff,
\newblock JHEP {\bf 02}, 044 (2002).
%%CITATION = HEP-PH 0109036;%%

\bibitem{Pukhov:2004ca}
A.~Pukhov,
\newblock (2004), [hep-ph/0412191].
%%CITATION = HEP-PH 0412191;%%

\bibitem{Pukhov:1999gg}
A.~Pukhov{ et~al.},
\newblock (1999), [hep-ph/9908288].
%%CITATION = HEP-PH 9908288;%%

\bibitem{Tanaka:1997qn}
Minami-Tateya collaboration, H.~Tanaka, M.~Kuroda, T.~Kaneko, M.~Jimbo, T.~Kon,
\newblock Nucl. Instrum. Meth. {\bf A389}, 295 (1997).
%%CITATION = NUIMA,A389,295;%%

\bibitem{Cho:2006sx}
G.~C. Cho{ et~al.},
\newblock (2006), [hep-ph/0601063].
%%CITATION = HEP-PH 0601063;%%

\bibitem{Moretti:2001zz}
M.~Moretti, T.~Ohl, J.~Reuter,
\newblock (2001), [hep-ph/0102195].
%%CITATION = HEP-PH 0102195;%%

\bibitem{Gleisberg:2003xi}
T.~Gleisberg{ et~al.},
\newblock JHEP {\bf 02}, 056 (2004).
%%CITATION = HEP-PH 0311263;%%

\bibitem{Tsuno:2005ih}
S.~Tsuno,
\newblock (2005), [hep-ph/0501174].
%%CITATION = HEP-PH 0501174;%%

\bibitem{Maltoni:2002qb}
F.~Maltoni, T.~Stelzer,
\newblock JHEP {\bf 02}, 044 (2002).
%%CITATION = HEP-PH 0109036;%%

\bibitem{Kilian:2001qz}
W.~Kilian,
\newblock LC-TOOL-2001-039.

\bibitem{Boos:2001cv}
E.~Boos{ et~al.},
\newblock (2001),
\newblock hep-ph/0109068.
%%CITATION = HEP-PH 0109068;%%

\bibitem{Corcella:2000bw}
G.~Corcella{ et~al.},
\newblock JHEP {\bf 01}, 010 (2001).
%%CITATION = HEP-PH 0011363;%%

\bibitem{Sjostrand:2000wi}
T.~Sj{\"o}strand{ et~al.},
\newblock Comput. Phys. Commun. {\bf 135}, 238 (2001),
\newblock (see also the manual, hep-ph/0308153).
%%CITATION = HEP-PH 0010017;%%

\bibitem{Dreiner:1999qz}
H.~K. Dreiner, P.~Richardson, M.~H. Seymour,
\newblock JHEP {\bf 04}, 008 (2000).
%%CITATION = HEP-PH 9912407;%%

\bibitem{Richardson:2000nt}
P.~Richardson,
\newblock (2000), [hep-ph/0101105].
%%CITATION = HEP-PH 0101105;%%

\bibitem{Skands:2001it}
P.~Z. Skands,
\newblock Eur. Phys. J. {\bf C23}, 173 (2002).
%%CITATION = HEP-PH 0110137;%%

\bibitem{Skands:2002qz}
P.~Z. Skands,
\newblock (2002), [hep-ph/0209199].
%%CITATION = HEP-PH 0209199;%%

\bibitem{Sjostrand:2002ip}
T.~Sj{\"o}strand, P.~Z. Skands,
\newblock Nucl. Phys. {\bf B659}, 243 (2003).
%%CITATION = HEP-PH 0212264;%%

\bibitem{Norrbin:2000uu}
E.~Norrbin, T.~Sj{\"o}strand,
\newblock Nucl. Phys. {\bf B603}, 297 (2001).
%%CITATION = HEP-PH 0010012;%%

\bibitem{Pukhov:2005je}
A.~Pukhov, P.~Skands,
\newblock Les {Houches} squared event generator for the {NMSSM},
\newblock FERMILAB-CONF-05-520-T. In {Les Houches} {'Physics at TeV Colliders
  2005'} {BSM Working Group}: {Summary report}, hep-ph/0602198.

\bibitem{Kraan:2004tz}
A.~C. Kraan,
\newblock Eur. Phys. J. {\bf C37}, 91 (2004).
%%CITATION = HEP-EX 0404001;%%

\bibitem{Kraan:2005ji}
A.~C. Kraan, J.~B. Hansen, P.~Nevski,
\newblock (2005),
\newblock {\\}(see {\tt http://www.thep.lu.se/$\sim$torbjorn/Pythia.html}),
  [hep-ex/0511014].
%%CITATION = HEP-EX 0511014;%%

\bibitem{Ghodbane:1999va}
N.~Ghodbane, S.~Katsanevas, P.~Morawitz, E.Perez,
\newblock (1999), [hep-ph/9909499].
%%CITATION = HEP-PH 9909499;%%

\bibitem{Bechtle:2004pc}
P.~Bechtle, K.~Desch, P.~Wienemann,
\newblock (2004),
\newblock hep-ph/0412012.
%%CITATION = HEP-PH 0412012;%%

\bibitem{Lafaye:2004cn}
R.~Lafaye, T.~Plehn, D.~Zerwas,
\newblock (2004),
\newblock hep-ph/0404282.
%%CITATION = HEP-PH 0404282;%%

\bibitem{Bardin:1999yd}
D.~Y. Bardin{ et~al.},
\newblock Comput. Phys. Commun. {\bf 133}, 229 (2001).
%%CITATION = HEP-PH 9908433;%%

\end{thebibliography}

\end{document}